\documentclass[12pt]{article}
\usepackage{axodraw}
\usepackage{pstricks}
\usepackage{color}

\usepackage{array}
\usepackage{epsfig}
\usepackage{amssymb}
\usepackage{graphics,graphpap}
\usepackage{amssymb}
\usepackage{amsmath}
\usepackage{slashed}
\usepackage{dsfont}

\def\nnb{\nonumber}
\def\nnb{\nonumber}
\newcommand{\f}{\frac}
\newcommand{\tev}{\, {\rm TeV}}

\newcommand{\Heff}{{\cal H}_\text{ eff}}

\newcommand{\be}{\begin{equation}}
\newcommand{\ee}{\end{equation}}
\newcommand{\bea}{\begin{eqnarray}}
\newcommand{\eea}{\end{eqnarray}}

\newcommand{\bi}{\begin{itemize}}
\newcommand{\ei}{\end{itemize}}
\newcommand{\ord}{{\cal O}}

\usepackage{graphicx}

\medskipamount 

\usepackage{fancyhdr}
\advance \headheight by 3.0truept       

\newlength{\textlength}
\newlength{\overlinelength}

\def\bbuildrel#1_#2^#3{\mathrel{\mathop{\kern 0pt#1}\limits_{#2}^{#3}}}
 \def\s#1{\setbox0=\hbox{$#1$}%
   \rlap{\ifdim\wd0>.7em\kern.22\wd0\else\kern.1\wd0\fi /}#1}


\begin{document}

\begin{titlepage}
\begin{flushright}
\begin{tabular}{l}
TUM-HEP-825/12\\
FLAVOUR(267104)-ERC-8
\end{tabular}
\end{flushright}
\vskip1.25cm
\begin{center}
{\Large \bf \boldmath
Completing NLO QCD Corrections  
for Tree Level Non-Leptonic
$\Delta F = 1$ Decays Beyond the 
Standard Model
}
\vskip1.3cm
{\bf
Andrzej J. Buras$^{a,b}$ and 
Jennifer Girrbach$^{b,c}$}
\vskip0.5cm
$^a$ Physik Department, Technische Universit{\"a}t M{\"u}nchen,
D-85748 Garching, Germany
\\
$^b$ TUM-IAS, Lichtenbergstr. 2a, D-85748 Garching, Germany\\
$^c$ Excellence Cluster Universe, TUM, Boltzmannstra\ss{}e 2, D-85748 Garching\\
\vskip0.1cm


\vskip0.1cm

{\large\bf Abstract\\[10pt]} \parbox[t]{\textwidth}{

In various extensions of the Standard Model (SM) 
tree level non-leptonic decays of 
hadrons receive contributions from new heavy gauge bosons and scalars. 
Prominent examples are the right-handed $W^\prime$ bosons in left-right symmetric models and charged Higgs ($H^\pm$) particles in models with extended scalar sector like  two Higgs doublet models and supersymmetric models. 
Even in the case of decays with four different quark flavours involved, 
to which penguin operators cannot contribute, twenty linearly independent 
operators, instead of two in the SM, have to be considered.
Anticipating the important role of such decays at the LHCb, KEKB and Super-B in 
Rome and having in mind future improved lattice computations, we complete 
the existing NLO QCD  formulae for these processes by  calculating
$\mathcal{O}(\alpha_s)$
corrections to matching conditions for the Wilson coefficients of all 
contributing operators in the NDR-$\overline{\text{MS}}$ scheme.
This
allows to reduce certain unphysical scale and renormalization scheme
dependences in the existing NLO calculations. Our results can also be 
applied to models with tree-level heavy neutral gauge boson and scalar 
exchanges in $\Delta F = 1$ transitions and constitute an important part of NLO analyses of those non-leptonic
 decays to which 
also penguin operators contribute. 
}

\vfill

\end{center}
\end{titlepage}

\setcounter{footnote}{0}

\newpage

\section{Introduction}

In the Standard Model (SM) the non-leptonic $\Delta F=1$ decays of mesons are 
governed by the $(V-A)\times (V-A)$ structure of the leading four-quark 
operators originating in the tree-level $W^\pm$ exchanges. If all the four 
flavours of the participating quarks are different from each other the only 
possible diagrams contributing to these decays in the SM and in any of its extensions are the current-current ones: penguin diagrams are absent. Decays of 
this type are theoretically cleaner than the ones in which also penguin 
diagrams and penguin operators contribute. As such they are well suited for 
the determination of the CKM parameters, in particular the angles $\gamma$ and 
$\beta$ in the unitarity triangle 
\cite{Fleischer:2011sg,Buchalla:2008tg,Antonelli:2009ws}.

While non-leptonic decays are subject to significant non-perturbative 
uncertainties originating in hadronic matrix elements of four-quark operators, the 
QCD factorization approach to non-leptonic two-body decays \cite{Beneke:1999br} 
combined with advanced lattice calculations could one day promote non-leptonic 
two-body meson decays to precise tools in testing the SM and its extensions. 
In these studies renormalization group short distance QCD effects play an 
important role. In the SM they are known including  the NLO corrections 
 and  in a few 
processes at the NNLO level. An up-to-date review can be found in 
\cite{Buras:2011we}.

Beyond the SM new local four-quark operators with different Dirac structures 
can be generated. The simplest example are $(V+A)\times(V+A)$ operators originating 
in the exchange of $W^{\pm\prime}$ gauge bosons in the left-right symmetric 
models. However, also right-handed (RH) couplings of the SM $W^\pm$ gauge 
bosons can be generated in various extensions of the SM like left-right 
symmetric models and generally also models with vectorial heavy quarks that mix 
with the SM chiral quarks. In this case also  $(V-A)\times(V+A)$ operators 
contribute. The latter operators generate through QCD corrections 
 $(S-P)\times(S+P)$ operators present also in models with charged ($H^\pm$) 
Higgs particles. In the latter models also $(S\pm P)\times(S\pm P)$ operators 
are present. Needless to say all these statements also apply to neutral 
gauge bosons and scalars mediating $\Delta F=1$ transitions.

The full set of twenty linearly independent 
dimension six four-quark operators with four different 
flavours in all extensions of the SM has been listed in 
\cite{Ciuchini:1997bw,Ciuchini:1998ix,Buras:2000if}, where also two-loop QCD 
anomalous dimensions of these operators have been calculated. However, the 
full NLO QCD renormalization group analysis of non-leptonic decays requires also the calculation of
$\mathcal{O}(\alpha_s)$
corrections to matching conditions for the Wilson coefficients of the 
operators in question. While such corrections are known within the SM 
\cite{Altarelli:1980fi,Buras:1989xd} at the NLO level, to our 
knowledge a complete analysis of these corrections including all operators in 
 any extension of the SM is absent in the 
literature. 

In a recent paper \cite{Buras:2012fs} we have calculated $\mathcal{O}(\alpha_s)$
corrections to matching conditions for the Wilson coefficients relevant for 
$\Delta F=2$ processes mediated by heavy colourless 
neutral gauge bosons and scalars reducing thereby certain unphysical scale and renormalization scheme dependences 
present in the absence of such corrections. Similar 
unphysical scale and renormalization scheme dependences are present in the absence of $\ord(\alpha_s)$ matching corrections in 
$\Delta F=1$ amplitudes generated by tree-level gauge boson and scalar 
exchanges and it is desirable to reduce them as well.

The main goal of our paper is the calculation of
$\mathcal{O}(\alpha_s)$
corrections to matching conditions for the Wilson coefficients of all 
dimension six four-quark operators with four different 
flavours contributing to $\Delta F=1$ decays mediated by colourless gauge 
bosons and scalars
in the NDR-$\overline{\text{MS}}$ scheme. As the two-loop anomalous dimensions 
for these operators have been already calculated in this scheme in 
\cite{Buras:2000if} our calculations complete the NLO QCD analysis of the
decays in question.

Our paper is organized as follows. In Section~\ref{Sec2} we recall the general structure 
of the effective Hamiltonians for $\Delta F=1$ processes in question 
and we give the full 
list of four-fermion operators that contribute to these transitions. 
Subsequently we collect their one- and two-loop anomalous dimension matrices. 
In 
Section~\ref{Sec3} we calculate  $\mathcal{O}(\alpha_s)$ corrections 
to  the amplitudes in the full theory and in Section~\ref{Sec4} the 
corresponding results for the matrix elements of operators are presented. 
This allows us in Section~\ref{Sec5} to present the Wilson coefficients 
of all twenty operators including  $\mathcal{O}(\alpha_s)$ corrections.
In Section~\ref{MasterNLO} combining our results with the known renormalization 
group evolution matrices we arrive at a complete NLO formulae for the 
Wilson coefficients of the involved operators.
In Section~\ref{Sec6} we demonstrate the scale independence of the 
physical amplitudes analytically and in Section~\ref{Sec6a} we investigate 
the removal of this scale dependence numerically.
We conclude with a brief summary in Section~\ref{Sec7}.

 \section{Theoretical Framework}\label{Sec2}
 \subsection{Local Operators}
While in the SM only two current-current operators contribute to each 
$\Delta F=1$ transition, 
the list of current-current operators beyond the SM is much longer.
As in \cite{Buras:2000if} 
we choose the operators in such a manner that all the
four flavours they contain are different. In such a case, 
the only possible diagrams are the
current--current ones.  In what follows we will fix the four flavours
to be ${b}$, $u$, ${c}$, $d$ but other choices are clearly 
possible without changing our results. 

Twenty linearly independent operators can be built out of four
different quark fields.  They can be split into eight separate sectors,
between which there is no mixing. The operators belonging to the first
two sectors (VLL, VLR), that are relevant for gauge boson 
contributions, are given as follows
\begin{subequations} \label{normal.df1}
\bea 
Q_1^{\rm VLL} &=& (\bar{b}^{\alpha} \gamma_{\mu}    P_L u^{ \beta})
              (\bar{c}^{ \beta} \gamma^{\mu}    P_L d^{\alpha}),
\\
Q_2^{\rm VLL} &=& (\bar{b}^{\alpha} \gamma_{\mu}    P_L u^{\alpha})
              (\bar{c}^{ \beta} \gamma^{\mu}    P_L d^{ \beta}),
\\[4mm]
Q_1^{\rm VLR} &=& (\bar{b}^{\alpha} \gamma_{\mu}    P_L u^{ \beta})
              (\bar{c}^{ \beta} \gamma^{\mu}    P_R d^{\alpha}),
\\
Q_2^{\rm VLR} &=& (\bar{b}^{\alpha} \gamma_{\mu}    P_L u^{\alpha})
              (\bar{c}^{ \beta} \gamma^{\mu}    P_R d^{ \beta}),
\eea
\end{subequations}
where $\alpha,\beta$ denote quark colours. In the case of scalar 
contributions the following operators have to be considered:
\begin{subequations} \label{normal.df1a}
\bea
Q_1^{\rm SLR} &=& (\bar{b}^{\alpha}                 P_L u^{ \beta})
              (\bar{c}^{ \beta}                 P_R d^{\alpha}),
\\
Q_2^{\rm SLR} &=& (\bar{b}^{\alpha}                 P_L u^{\alpha})
              (\bar{c}^{ \beta}                 P_R d^{ \beta}),    
\eea
\end{subequations}
\begin{subequations}\label{normal.df1b}
\bea
Q_1^{\rm SLL} &=& (\bar{b}^{\alpha}                 P_L u^{ \beta})
              (\bar{c}^{ \beta}                 P_L d^{\alpha}),
\\
Q_2^{\rm SLL} &=& (\bar{b}^{\alpha}                 P_L u^{\alpha})
              (\bar{c}^{ \beta}                 P_L d^{\beta}),
\\
Q_3^{\rm SLL} &=& (\bar{b}^{\alpha} \sigma_{\mu\nu} P_L u^{ \beta})
              (\bar{c}^{ \beta} \sigma^{\mu\nu} P_L d^{\alpha}),
\\
Q_4^{\rm SLL} &=& (\bar{b}^{\alpha} \sigma_{\mu\nu} P_L u^{\alpha})
              (\bar{c}^{ \beta} \sigma^{\mu\nu} P_L d^{ \beta}).
\eea
\end{subequations}

The operators belonging to the four remaining sectors (VRR, VRL, SRL
and SRR) are obtained from the above by interchanging $P_L$ and $P_R$.
Obviously, it is sufficient to calculate the Wilson coefficients 
only for the VLL,
VLR, SLR and SLL sectors.  The ``mirror'' operators in the VRR, VRL,
SRL and SRR sectors will have exactly the same properties under QCD
renormalization.

The two-loop anomalous dimensions for these operators have been calculated
in the  NDR-$\overline{\text{MS}}$ scheme in
\cite{Buras:2000if} with a particular choice of the evanescent operators.
As discussed there, while these operators are essential for the correct
evaluation of two-loop matrix elements, 
the virtue of the formulation of the 
 NDR-$\overline{\text{MS}}$ scheme introduced in \cite{Buras:1989xd} is that 
 the evanescent operators defined in this scheme 
influence only two-loop anomalous dimensions. By definition they do not 
contribute to the matching and to the finite gluon corrections to the matrix elements  of renormalized operators calculated by us. They are
simply subtracted away in
the process of renormalization. This issue is summarized in Section 6.9.4 of 
\cite{Buras:1998raa}, where further references can be found. 
A very important paper in this respect is also the one of Herrlich and Nierste 
\cite{Herrlich:1994kh}, where this issue is discussed in full generality. 
Therefore effectively the  one-loop
calculations presented here and in \cite{Buras:2012fs} are  based on the projections 
of various Dirac structures on physical operators that are 
consistent with  \cite{Buras:2000if,Buras:1989xd} but otherwise the evanescent 
operators can be dropped from the beginning.

\subsection{Renormalization Group Functions}
\subsubsection{Running of QCD Coupling and Running Masses}
For the complete NLO renormalization group analysis we need a number 
of renormalization group functions that we recall here for completeness.

In particular we need the QCD $\beta$ function at the two-loop level
\begin{equation}\label{bg01}
\beta(g)=-\beta_0\frac{g^3}{16\pi^2}-\beta_1\frac{g^5}{(16\pi^2)^2},  
\end{equation}
where
\begin{equation}\label{b0b1}
\beta_0=\frac{11N-2f}{3}\,,\qquad
\beta_1=\frac{34}{3}N^2-\frac{10}{3}Nf-2C_F f\,,\qquad
C_F=\frac{N^2-1}{2N}\,.
\end{equation}
$f$ is number of flavours and $N$ the number of colours.

Similarly, the two-loop expression for the quark mass anomalous
dimension can be written as
\begin{equation}\label{gama}
\gamma_m(\alpha_s)=\gamma^{(0)}_{m}\frac{\alpha_s}{4\pi}
 + \gamma^{(1)}_{m}\left( \frac{\alpha_s}{4\pi}\right)^2,
\end{equation}
where
\begin{equation}
\label{gm01} 
\gamma^{(0)}_{m}=6C_F\qquad \gamma^{(1)}_{m}
=C_F\left(3C_F+\frac{97}{3}N-\frac{10}{3}f\right).  
\end{equation}

\subsubsection{One-Loop and Two-Loop Anomalous Dimension Matrices of Operators}

The most important ingredients of any renormalization group analysis in weak decays are the anomalous dimension matrices that we define in general form as 
follows
\begin{equation}\label{gamaoperator}
\hat{\gamma}(\alpha_s)=\hat{\gamma}^{(0)}\frac{\alpha_s}{4\pi}
 + \hat{\gamma}^{(1)}\left( \frac{\alpha_s}{4\pi}\right)^2.
\end{equation}
In particular the one-loop anomalous dimension matrices $\hat{\gamma}^{(0)}$ 
will play important role in our discussion of
the removal of the unphysical $\mu$ dependences at NLO in Section~\ref{Sec6}. 

The one-loop matrices for all operators considered in this paper read
{\allowdisplaybreaks
\bea
\hat{\gamma}^{(0)\rm VLL} &=& \left( \begin{array}{ccc}
-\f{6}{N} && 6 \\[1mm] 
6 && -\f{6}{N}
\end{array} \right), \label{ga0VLL} \\[2mm]
\hat{\gamma}^{(0)\rm VLR} &=& \left( \begin{array}{ccc}
- 6N +\f{6}{N} && 0 \\[1mm] 
-6 && \f{6}{N}
\end{array} \right), \label{ga0VLR} \\[2mm]
\hat{\gamma}^{(0)\rm SLR} &=& \left( \begin{array}{ccc}
\f{6}{N} && -6 \\[1mm] 
0 && - 6N+ \f{6}{N} 
\end{array} \right), \label{ga0SLR} \\[2mm]
\hat{\gamma}^{(0)\rm SLL} &=& \left( \begin{array}{cccc}
\f{6}{N} & -6 & \f{N}{2}-\f{1}{N} & \f{1}{2} \\ 
0 & - 6N+\f{6}{N} & 1 & -\f{1}{N} \\ 
-\f{48}{N} + 24N & 24 & -\f{2}{N} - 4N & 6 \\ 
48  & -\f{48}{N} & 0 & 2N-\f{2}{N} 
\end{array} \right)\,.\label{ga0SLL}
\eea}%
The two-loop matrices for all operators  in the  NDR-$\overline{\text{MS}}$ scheme  used in this paper read \cite{Buras:2000if} 
\bea
\hat{\gamma}^{(1)\rm VLL} &=& \left( \begin{array}{ccc}
-\f{22}{3} - \f{57}{2N^2} - \f{2}{3N} f &~~&
- \f{19}{6} N + \f{39}{N} +\f{2}{3}f \\[1mm] 
- \f{19}{6} N + \f{39}{N} +\f{2}{3}f &&
-\f{22}{3} - \f{57}{2N^2} - \f{2}{3N} f
\end{array} \right), \label{ga1VLL}\\[2mm]
\hat{\gamma}^{(1)\rm VLR} &=& \left( \begin{array}{ccc}
-\f{203}{6} N^2 +\f{479}{6} + \f{15}{2N^2} + \f{10}{3} N f -\f{22}{3N} f &~~~&
-\f{71}{2} N - \f{18}{N} +4f \\[1mm] 
- \f{100}{3} N + \f{3}{N} + \f{22}{3} f &&
\f{137}{6} + \f{15}{2N^2} - \f{22}{3N} f
\end{array} \right), \label{ga1VLR} \\[2mm]
\hat{\gamma}^{(1)\rm SLR} &=& \left( \begin{array}{ccc}
\f{137}{6} + \f{15}{2N^2} - \f{22}{3N} f &~~~&
-\f{100}{3} N + \f{3}{N} + \f{22}{3} f \\[1mm] 
- \f{71}{2} N - \f{18}{N} + 4f &&
-\f{203}{6} N^2 +\f{479}{6} + \f{15}{2N^2} + \f{10}{3} N f -\f{22}{3N} f
\end{array} \right), \label{ga1SLR} 
\eea
{\allowdisplaybreaks
\bea 
\gamma^{(1)\rm SLL}_{11} &=& \begin{array}{l}
-\f{N^2}{2} + \f{148}{3} - \f{107}{2N^2} - 2 N f - \f{10}{3N} f, 
\end{array}\nnb\\
\gamma^{(1)\rm SLL}_{12} &=& \begin{array}{l}
-\f{178}{3} N + \f{64}{N} + \f{16}{3} f, 
\end{array}\nnb\\
\gamma^{(1)\rm SLL}_{13} &=& \begin{array}{l}
\f{107}{36} N^2 -\f{71}{18} - \f{4}{N^2} - \f{1}{18} N f + \f{f}{9N}, 
\end{array}\nnb\\
\gamma^{(1)\rm SLL}_{14} &=& \begin{array}{l}
-\f{109}{36} N + \f{8}{N} -\f{f}{18},
\end{array}\nnb\\
\gamma^{(1)\rm SLL}_{21} &=& \begin{array}{l}
-26N + \f{104}{N}, 
\end{array}\nnb\\
\gamma^{(1)\rm SLL}_{22} &=& \begin{array}{l}
-\f{203}{6} N^2 + \f{28}{3} - \f{107}{2N^2} + \f{10}{3} N f - \f{10}{3N} f, 
\end{array} \nonumber\\
\gamma^{(1)\rm SLL}_{23} &=& \begin{array}{l}
\f{89}{18} N + \f{2}{N} -\f{1}{9} f, 
\end{array}\nnb\\
\gamma^{(1)\rm SLL}_{24} &=& \begin{array}{l}
-\f{53}{18} - \f{4}{N^2} + \f{1}{9N} f, 
\end{array}\nnb\\
\gamma^{(1)\rm SLL}_{31} &=& \begin{array}{l}
\f{676}{3} N^2 -\f{1880}{3} - \f{320}{N^2} - \f{88}{3} N f + \f{176}{3N} f, 
\end{array}\nnb\\
\gamma^{(1)\rm SLL}_{32} &=& \begin{array}{l}
\f{820}{3} N + \f{448}{N} -\f{88}{3} f, 
\end{array}\nnb\\
\gamma^{(1)\rm SLL}_{33} &=& \begin{array}{l}
-\f{257}{18} N^2 -\f{116}{9} + \f{21}{2N^2} + \f{22}{9} N f + \f{2}{9N} f, 
\end{array}\nnb\\
\gamma^{(1)\rm SLL}_{34} &=& \begin{array}{l}
\f{50}{3} N -\f{8}{3} f, 
\end{array}\nnb\\
\gamma^{(1)\rm SLL}_{41} &=& \begin{array}{l}
\f{488}{3} N + \f{416}{N} -\f{176}{3} f, 
\end{array}\nnb\\
\gamma^{(1)\rm SLL}_{42} &=& \begin{array}{l}
-\f{776}{3} - \f{320}{N^2} + \f{176}{3N} f, 
\end{array}\nnb\\
\gamma^{(1)\rm SLL}_{43} &=& \begin{array}{l}
\f{22}{3} N  - \f{40}{N}  +\f{8}{3} f, 
\end{array}\nnb\\
\gamma^{(1)\rm SLL}_{44} &=& \begin{array}{l}
\f{343}{18} N^2 + \f{28}{9} + \f{21}{2N^2} - \f{26}{9} N f + \f{2}{9N} f. 
\end{array}\label{ga1bSLL} 
\eea}%

\subsection{Effective Hamiltonian}

The effective Hamiltonian for $\Delta F=1$ transitions can be written 
in a general form as follows\footnote{In what follows we drop for simplicity $h.c.$}
\be\label{Heff-general}
\Heff^{\Delta F=1} =\kappa
\sum_{i,a}C^a_i(\mu)Q_i^a\,,
\ee
where $Q_i^a$ are the operators given in Eqs.~(\ref{normal.df1}), (\ref{normal.df1a}) and (\ref{normal.df1b}) and
$C_i^a(\mu)$ their Wilson coefficients evaluated at a scale $\mu$ at which 
the hadronic matrix elements are evaluated. 
The scale $\mu$ can be the low 
energy scale $\mu_L$ at which actual lattice calculations 
are performed or any other 
scale, in particular the matching scale $\mu_\text{in}$. 
In this case the matrix elements are obtained by evolving lattice results by means of
renormalization group (RG) equations  from $\mu_L$ to $\mu_\text{in}$. The 
resulting matrix elements of the effective Hamiltonian, that are directly 
related to decay amplitudes, can then be 
written generally as follows:
\be\label{Heff-general1}
\langle\Heff^{\Delta F=1}\rangle =\kappa
\sum_{i,a}C^a_i(\mu_\text{in})\langle Q_i^a(\mu_\text{in})\rangle\,.
\ee
The overall factor $\kappa$  in our analysis depends on the exchanged 
boson (vector or scalar) 
 and will be chosen such that for non-vanishing Wilson coefficients 
$C_i^a(\mu_{\rm in}) = 1$ in  LO. 

Evidently the matrix elements $\langle Q_i^a(\mu_\text{in})\rangle$ depend 
on the matching scale $\mu_\text{in}$ and also on the renormalization 
scheme for operators. In order to remove these unphysical dependences from physical amplitudes one 
has to calculate $\mathcal{O}(\alpha_s)$ corrections to $C^a_i(\mu_\text{in})$. 
This is the goal of Sections~\ref{Sec3}-\ref{Sec5}.

\subsection{Procedure for Matching}

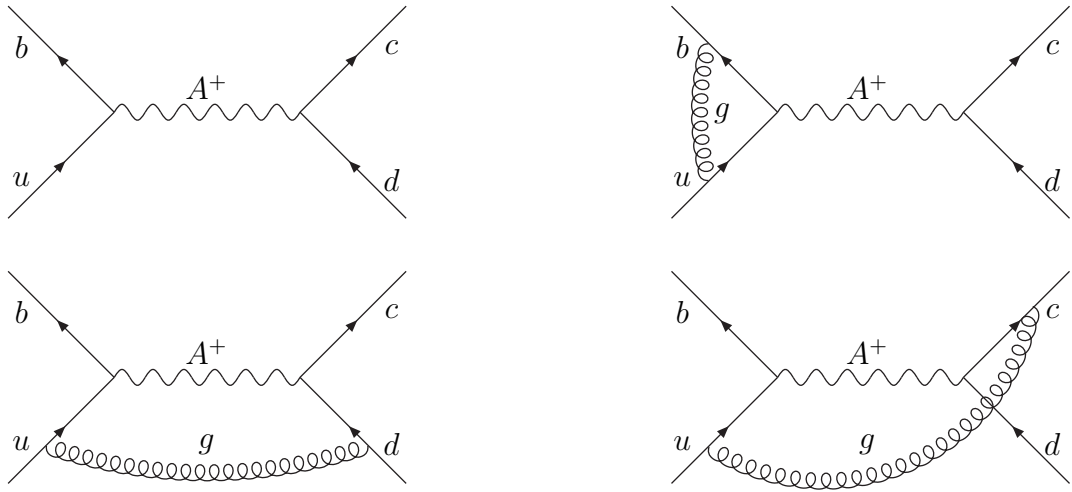
\begin{figure}[!tbh]
\begin{center}
\scalebox{1.0}{
    \begin{picture}(400,200)(0,0)
      \SetColor{Black}

	\ArrowLine(40,150)(0,190)\Text(5,175)[]{$ b$}
	\ArrowLine(0,110)(40,150)\Text(5,125)[]{$u$}
	\Photon(40,150)(110,150){3}{6}\Text(75,160)[c]{$A^+$}
	\ArrowLine(110,150)(150,190)\Text(145,175)[]{$ c$}
	\ArrowLine(150,110)(110,150)\Text(145,125)[]{$  d$}
	
	\ArrowLine(290,150)(250,190)\Text(255,175)[]{$ b$}
	\ArrowLine(250,110)(290,150)\Text(255,125)[]{$u$}
	\Photon(290,150)(360,150){3}{6}\Text(325,160)[c]{$A^+$}
	\ArrowLine(360,150)(400,190)\Text(395,175)[]{$ c$}
	\ArrowLine(400,110)(360,150)\Text(395,125)[]{$  d$}
	\GlueArc(360,150)(99.45,165,195){3}{9}\Text(270,150)[]{$g$}

	\ArrowLine(40,50)(0,90)\Text(5,75)[]{$ b$}
	\ArrowLine(0,10)(40,50)\Text(5,25)[]{$u$}
	\Photon(40,50)(110,50){3}{6}\Text(75,60)[c]{$A^+$}
	\ArrowLine(110,50)(150,90)\Text(145,75)[]{$ c$}
	\ArrowLine(150,10)(110,50)\Text(145,25)[]{$  d$}
	\GlueArc(75,200)(186,-109,-71){3}{22}\Text(75,25)[c]{$g$}

	\ArrowLine(290,50)(250,90)\Text(255,75)[]{$b$}
	\ArrowLine(250,10)(290,50)\Text(255,25)[]{$u$}
	\Photon(290,50)(360,50){3}{6}\Text(325,60)[c]{$A^+$}
	\ArrowLine(360,50)(400,90)\Text(395,75)[]{$ c$}
	\ArrowLine(400,10)(360,50)\Text(395,25)[]{$  d$}
	\GlueArc(307.6,90.6)(79.75,-123,-10){3}{26}\Text(325,25)[c]{$g$}

         \end{picture}}
\caption{ \it Tree level diagram and one-loop QCD corrections to $\Delta B = 1$ transition mediated by a gauge
boson in the full theory. Mirror diagrams are not shown.}\label{fig:fullgauge}
\end{center}
\end{figure}

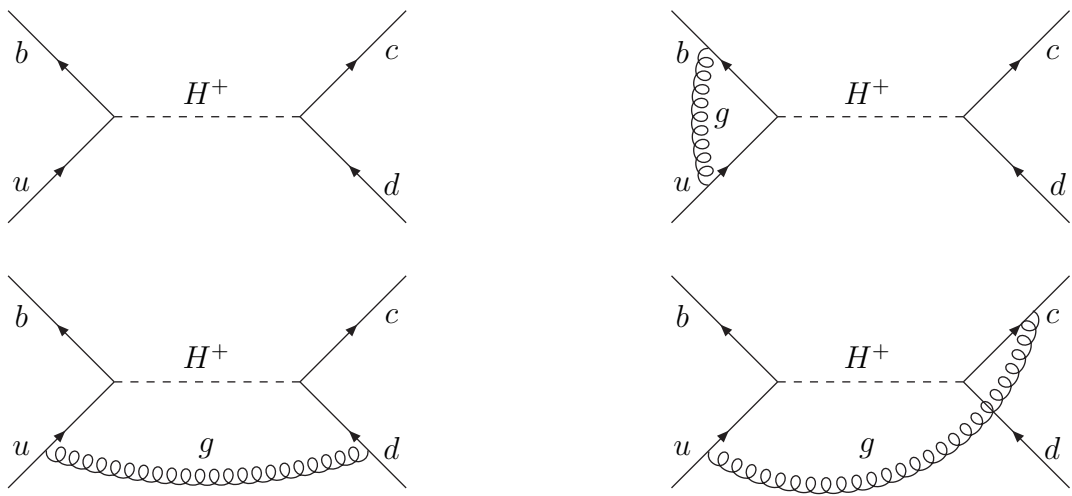
\begin{figure}[!tbh]
\begin{center}
\scalebox{1.0}{
    \begin{picture}(400,200)(0,0)
      \SetColor{Black}

	\ArrowLine(40,150)(0,190)\Text(5,175)[]{$b$}
	\ArrowLine(0,110)(40,150)\Text(5,125)[]{$u$}
	\DashLine(40,150)(110,150){3}\Text(75,160)[c]{$H^+$}
	\ArrowLine(110,150)(150,190)\Text(145,175)[]{$ c$}
	\ArrowLine(150,110)(110,150)\Text(145,125)[]{$  d$}

	\ArrowLine(290,150)(250,190)\Text(255,175)[]{$ b$}
	\ArrowLine(250,110)(290,150)\Text(255,125)[]{$u$}
	\DashLine(290,150)(360,150){3}\Text(325,160)[c]{$H^+$}
	\ArrowLine(360,150)(400,190)\Text(395,175)[]{$ c$}
	\ArrowLine(400,110)(360,150)\Text(395,125)[]{$ \ d$}
	\GlueArc(360,150)(99.45,165,195){3}{9}\Text(270,150)[]{$g$}

	\ArrowLine(40,50)(0,90)\Text(5,75)[]{$ b$}
	\ArrowLine(0,10)(40,50)\Text(5,25)[]{$u$}
	\DashLine(40,50)(110,50){3}\Text(75,60)[c]{$H^+$}
	\ArrowLine(110,50)(150,90)\Text(145,75)[]{$ c$}
	\ArrowLine(150,10)(110,50)\Text(145,25)[]{$  d$}
	\GlueArc(75,200)(186,-109,-71){3}{22}\Text(75,25)[c]{$g$}

	\ArrowLine(290,50)(250,90)\Text(255,75)[]{$ b$}
	\ArrowLine(250,10)(290,50)\Text(255,25)[]{$u$}
	\DashLine(290,50)(360,50){3}\Text(325,60)[c]{$H^+$}
	\ArrowLine(360,50)(400,90)\Text(395,75)[]{$ c$}
	\ArrowLine(400,10)(360,50)\Text(395,25)[]{$  d$}
	\GlueArc(307.6,90.6)(79.75,-123,-10){3}{26}\Text(325,25)[c]{$g$}

         \end{picture}}
\caption{\it Tree level diagram and one-loop QCD corrections to $\Delta B = 1$ transition mediated by a scalar
particle in the full theory. Mirror diagrams are not shown.}\label{fig:fullscalar}
\end{center}
\end{figure}

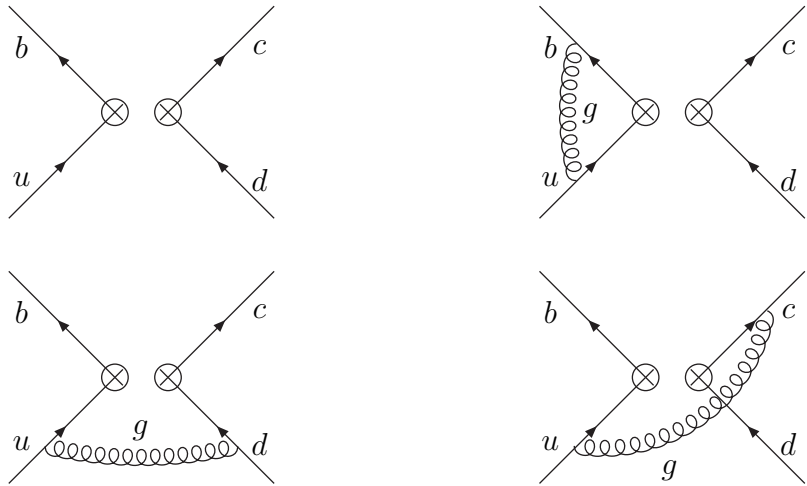
\begin{figure}[!tbh]
\begin{center}
\scalebox{1.0}{
    \begin{picture}(400,200)(0,0)
      \SetColor{Black}

	\ArrowLine(90,150)(50,190)\Text(55,175)[]{$ b$}
	\ArrowLine(50,110)(90,150)\Text(55,125)[]{$u$}
	\CArc(90,150)(5,0,360)  
	\Line(90,150)(93,153)\Line(90,150)(93,147)

	\ArrowLine(110,150)(150,190)\Text(145,175)[]{$ c$}
	\ArrowLine(150,110)(110,150)\Text(145,125)[]{$  d$}
	\CArc(110,150)(5,0,360) 
	\Line(110,150)(107,153)\Line(110,150)(107,147)

	\ArrowLine(290,150)(250,190)\Text(255,175)[]{$ b$}
	\ArrowLine(250,110)(290,150)\Text(255,125)[]{$u$}
	\CArc(290,150)(5,0,360) 
	 \Line(290,150)(293,153)\Line(290,150)(293,147)
\GlueArc(360,150)(99.45,165,195){3}{9}\Text(270,150)[]{$g$}
	\ArrowLine(310,150)(350,190)\Text(345,175)[]{$ c$}
	\ArrowLine(350,110)(310,150)\Text(345,125)[]{$  d$}
	\CArc(310,150)(5,0,360) 
	\Line(310,150)(307,153)\Line(310,150)(307,147)

	\ArrowLine(90,50)(50,90)\Text(55,75)[]{$ b$}
	\ArrowLine(50,10)(90,50)\Text(55,25)[]{$u$}
	\CArc(90,50)(5,0,360)  
	\Line(90,50)(93,53)\Line(90,50)(93,47)
      \GlueArc(100,175)(155.23,-103.5,-76.5){3}{13}\Text(100,30)[]{$g$}
	\ArrowLine(110,50)(150,90)\Text(145,75)[]{$ c$}
	\ArrowLine(150,10)(110,50)\Text(145,25)[]{$  d$}
	\CArc(110,50)(5,0,360) 
	\Line(110,50)(107,53)\Line(110,50)(107,47)

	\ArrowLine(290,50)(250,90)\Text(255,75)[]{$ b$}
	\ArrowLine(250,10)(290,50)\Text(255,25)[]{$u$}
	\CArc(290,50)(5,0,360) 
	 \Line(290,50)(293,53)\Line(290,50)(293,47)
 \GlueArc(274,86)(62.8,-100,-10){3}{16}\Text(300,15)[]{$g$}
	\ArrowLine(310,50)(350,90)\Text(345,75)[]{$ c$}
	\ArrowLine(350,10)(310,50)\Text(345,25)[]{$  d$}
	\CArc(310,50)(5,0,360) 
	\Line(310,50)(307,53)\Line(310,50)(307,47)

         \end{picture}}
\caption{\it Leading order and one-loop diagrams in the effective theory. 
Mirror diagrams are not shown.}\label{fig:eff}
\end{center}
\end{figure}

Before entering the details let us recall that
the calculations of $\ord(\alpha_s)$ QCD corrections to Wilson coefficients 
are by now standard and have been described in several papers. 
In particular in \cite{Buras:1990fn,Buras:1998raa} in the case of the SM. 
In \cite{Buras:2012fs} this procedure has been used to calculate  $\ord(\alpha_s)$ corrections to the Wilson 
coefficients of operators contributing to $\Delta F=2$ FCNC processes 
mediated by tree level neutral gauge  boson and scalar exchanges. Here we 
recall briefly this procedure that we will use in the case of $\Delta F=1$ 
decays.
 
\subsubsection*{Step 1}

We first calculated the amplitudes in the full theory. This amounts to the 
calculation of the diagrams  
in Figs.~\ref{fig:fullgauge} and \ref{fig:fullscalar}, in the case of a gauge boson exchange and a scalar exchange,
respectively.
In the presence of massless gluons one encounters infrared divergences. 
We have regulated these 
divergences by a common external momentum $p$ with $-p^2>0$ for all external 
massless fields as we did in \cite{Buras:2012fs}. 
The ultraviolet 
divergences present in diagrams in 
 Fig.~\ref{fig:fullgauge} and \ref{fig:fullscalar} with gluon 
 corrections to vertices
have been regulated 
using dimensional regularization with anti-commuting $\gamma_5$ in $4-2\epsilon$
dimensions.

\subsubsection*{Step 2}

We have calculated the matrix elements of contributing operators by 
evaluating the diagrams in Fig.~\ref{fig:eff} making the same
assumptions about the 
external fields as in the first step. In contrast to step 1 one 
has to renormalize the operators. This we do in the $\overline{\text{MS}}$ 
renormalization scheme with anti-commuting $\gamma_5$, which corresponds 
to the NDR scheme used also in \cite{Buras:2000if} and \cite{Buras:2001ra}.

\subsubsection*{Step 3}

 We finally inserted the results of the two steps above into a formula 
 like the one in Eq.~(\ref{Heff-general1}) and comparing the coefficients of 
 operators appearing on the l.h.s (full theory) and r.h.s (effective 
 theory) we found the coefficients $C_i^a(\mu_{\rm in})$. As these 
 coefficients cannot depend on the infrared behaviour of the theory, 
 the dependences on $p^2$  found in the first two 
 steps have to cancel each other in the evaluation of   $C_i^a(\mu_{\rm in})$. 
 Indeed we verified this explicitly. The interested reader can do this 
as well by inspecting our intermediate results that we present in 
Sections~\ref{Sec3} and \ref{Sec4}, respectively.

 Very often in analyses of  new physics contributions the overall factor in front of the sum in Eq.~(\ref{Heff-general1}) is 
 chosen as in the SM. However, in our analysis it will be more convenient 
to use 
 in each case the normalization in which the Wilson coefficient of the
leading operator evaluated at the matching scale is 
 equal to unity in the absence of QCD corrections. In this manner the 
applications of our formulae in various new physics (NP) models will be facilitated.

After this preparation we are ready to present our calculations in three 
steps in question.

\section{Amplitudes in the Full Theory}\label{Sec3}

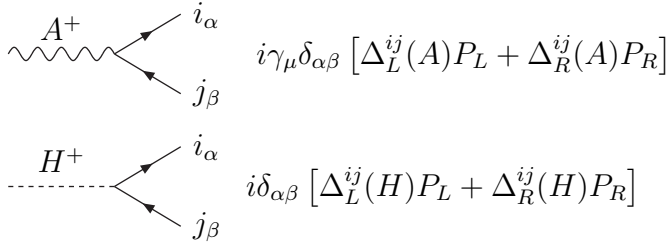
\begin{figure}[!htb]
\scalebox{1.0}{
    \begin{picture}(100,150)(0,-50)
      \SetColor{Black}

	\Photon(0,70)(40,70){2.5}{4.5}\ArrowLine(40,70)(65,85)\ArrowLine(65,55)(40,70) 
	\Text(20,80)[c]{$A^+$}\Text(70,85)[l]{$i_\alpha$}\Text(70,55)[l]{$j_\beta$}\Text(90,70)[l]{
$i\gamma_\mu\delta_{\alpha\beta}\left[
\Delta_L^{ij}(A)P_L+\Delta_R^{ij}(A)P_R\right]$}

	\DashLine(0,20)(40,20){2}\ArrowLine(40,20)(65,35)\ArrowLine(65,5)(40,20)  
	\Text(20,30)[c]{$H^+$}\Text(70,35)[l]{$i_\alpha$}\Text(70,5)[l]{$j_\beta$}\Text(90,20)[l]{$i\delta_{
\alpha\beta}\left[ \Delta_L^{ ij}
(H)P_L+\Delta_R^{ij}(H)P_R\right]$}
        \end{picture}}
\vskip-1.5cm
\caption{\it Feynman rules for colourless charged gauge boson $A^+$ with mass $M_A$, 
and charged colourless scalar particle $H^+$
with mass $M_H$, where $i\,(j)$ denotes an up-type (down-type) quark flavour with charge $+\frac{2}{3}$ ($-\frac{1}{3}$)  and
$\alpha,\,\beta$  are colour indices.}\label{fig:Feynman}
\end{figure}

Using the  Feynman rules in Fig.~\ref{fig:Feynman} we find for the 
colourless gauge boson exchange (after quark wave function renormalization)

\begin{align}
\begin{split}
 \mathcal{A}^\text{VLL} = & 
\frac{\left(\Delta_L^{ub}(A)\right)^\star\Delta_L^{cd}(A)}{M_A^2}\left[ \left(1+ 2
C_F\frac{\alpha_s}{4\pi}
\log\frac{\mu^2}{-p^2}\right) Q_2^\text{VLL} \right.\\
& \qquad\qquad\qquad\quad \left. +  \frac{\alpha_s}{4\pi}\left(\frac{1}{2}+
\log\frac{M^2_A}{-p^2}\right)\left[\frac{3}{N} Q_2^\text{VLL} -3 Q_1^\text{VLL}\right] \right]\end{split}
\\
\begin{split}
 \mathcal{A}^\text{VLR} =&  \frac{\left(\Delta_L^{ub}(A)\right)^\star\Delta_R^{cd}(A)}{M_A^2}\left[\left(1+ 2
C_F\frac{\alpha_s}{4\pi}
\log\frac{\mu^2}{-p^2} \right)Q_2^\text{VLR}\right.\\
& \qquad\qquad\qquad\quad\left. +  \frac{\alpha_s}{4\pi}\left(\frac{1}{2}+
\log\frac{M^2_A}{-p^2}\right)\left[-\frac{3}{N}Q_2^\text{VLR}+3 Q_1^\text{VLR}\right] \right]
\end{split}
\end{align}

For the colourless scalar exchange (after quark wave function and quark mass renormalizations) we find
\begin{align}
 \mathcal{A}^\text{SLR} = & -\frac{\left(\Delta_L^{ub}(H)\right)^\star\Delta_R^{cd}(H)}{M_H^2}\left[1 +
8 C_F \frac{\alpha_s}{4\pi} \left( 1+ \log\frac{\mu^2}{-p^2}\right)  \right]Q_2^\text{SLR}\\
\begin{split}
\mathcal{A}^\text{SLL}  = &-\frac{\left(\Delta_L^{ub}(H)\right)^\star\Delta_L^{cd}(H)}{2 M_H^2}\left[ \left(1 + 8
C_F\frac{\alpha_s}{4\pi}\left(1+
\log\frac{\mu^2}{-p^2}\right) \right)Q_2^\text{SLL}\right.\\
&\qquad\qquad\qquad\quad\left.+ \frac{\alpha_s}{4\pi} \left(\frac{1}{2}+\log\frac{M^2_H}{-p^2}\right)\left[\frac{1}{2N}Q_4^\text{SLL}
-\frac{1}{2} Q_3^\text{SLL} \right] \right]
\end{split}
\end{align}

\section{Matrix Elements of Operators}\label{Sec4}
 
Calculating the diagrams in Fig.~\ref{fig:eff} we find 
after quark wave function renormalization and operator renormalization 

{\allowdisplaybreaks
\begin{align}
 \langle Q_1^\text{VLL}\rangle &= \left[ 1 + 2 C_F \frac{\alpha_s}{4\pi}
\log\frac{\mu^2}{-p^2}\right] Q_1^\text{VLL}  + \frac{\alpha_s}{4
\pi} \left( \log\frac{\mu^2}{-p^2}+\frac{7}{3}\right)\left[\frac{3}{N} Q_1^\text{VLL} - 3 Q_2^\text{VLL}\right]\\
 \langle Q_2^\text{VLL}\rangle &= \left[ 1 + 2 C_F \frac{\alpha_s}{4\pi}
\log\frac{\mu^2}{-p^2}\right] Q_2^\text{VLL}  + \frac{\alpha_s}{4
\pi} \left( \log\frac{\mu^2}{-p^2}+\frac{7}{3}\right)\left[\frac{3}{N} Q_2^\text{VLL} - 3 Q_1^\text{VLL}\right]\\
 \langle Q_1^\text{VLR}\rangle &=\left[1+8 C_F \frac{\alpha_s}{4\pi}
\left(1+
\log\frac{\mu^2}{-p^2}\right) \right]Q_1^\text{VLR}   +\frac{\alpha_s}{4\pi}\left[\frac{3}{N}Q_1^\text{VLR}-
3Q_2^\text{VLR}\right]
\\
\begin{split}
 \langle Q_2^\text{VLR}\rangle &=\left[1+2 C_F \frac{\alpha_s}{4\pi}\left(1+
\log\frac{\mu^2}{-p^2}\right) \right]Q_2^\text{VLR} \\
& \qquad +\frac{\alpha_s}{4\pi}\left(\frac{1}{3} +
\log\frac{\mu^2}{-p^2}\right)\left[-\frac{3}{N}Q_2^\text{VLR}+3 Q_1^\text{VLR}\right]
\end{split}
\\
 \langle Q_1^\text{SLR}\rangle &=\left[1+2 C_F \frac{\alpha_s}{4\pi}
\log\frac{\mu^2}{-p^2} \right]Q_1^\text{SLR}   +\frac{\alpha_s}{4\pi}\left(\frac{1}{3} +
\log\frac{\mu^2}{-p^2}\right)\left[-\frac{3}{N}Q_1^\text{SLR}+
3Q_2^\text{SLR}\right]
\\
 \langle Q_2^\text{SLR}\rangle &=\left[1+8 C_F \frac{\alpha_s}{4\pi}\left(1+
\log\frac{\mu^2}{-p^2}\right) \right]Q_2^\text{SLR} +\frac{\alpha_s}{4\pi}\left[\frac{3}{N}Q_2^\text{VLR}-3 Q_1^\text{VLR}\right]
\\
  \langle Q_1^\text{SLL}\rangle &=\left[1+2 C_F \frac{\alpha_s}{4\pi}\left(\frac
{3}{2}+\log\frac{\mu^2}{-p^2}\right) \right]Q_1^\text{SLL}+
\frac{\alpha_s}{4\pi}\left(\frac{5}{6}+\log\frac{\mu^2}{-p^2}\right)\left[-\frac{3}{N} Q_1^\text{SLL}+ 3
Q_2^\text{SLL}\right]\nonumber\\
&\quad +\frac{\alpha_s}{4\pi} \left(2+\log\frac{\mu^2}{-p^2}\right) \left[\frac{2-N^2}{4N} Q_3^\text{SLL} -
\frac{1}{4}Q_4^\text{SLL}\right]\\
\begin{split}
  \langle Q_2^\text{SLL}\rangle &=\left[1+8 C_F \frac{\alpha_s}{4\pi}\left(1+\log\frac{\mu^2}{-p^2}\right)
\right]Q_2^\text{SLL}\\
&\qquad+\frac{\alpha_s}{4\pi} \left(2+\log\frac{\mu^2}{-p^2}\right)\left[\frac{1}{2N} Q_4^\text{SLL} -
\frac{1}{2}Q_3^\text{SLL}\right]
\end{split}\\
\begin{split}
 \langle Q_3^\text{SLL}\rangle & = \left[ 1 + \frac{\alpha_s}{4\pi} 3 N\left(\log\frac{\mu^2}{-p^2} +
\frac{5N^2 + 4}{6N^2}\right)\right] Q_3^\text{SLL} - \frac{\alpha_s}{4\pi}3\left(\frac{3}{2}+\log\frac{\mu^2}{-p^2}\right)Q_4^\text{SLL}
 \\
& \quad +\frac{\alpha_s}{4\pi} 12\left(\frac{2}{N}-N\right) \left(\log\frac{\mu^2}{-p^2} +
\frac{1}{3}\right)Q_1^\text{SLL} - \frac{\alpha_s}{4\pi}12\left(\frac{1}{3}+\log\frac{\mu^2}{-p^2}\right) Q_2^\text{SLL}
\end{split}\\
\begin{split}
\langle Q_4^\text{SLL}\rangle & = Q_4^\text{SLL} + 24
\frac{\alpha_s}{4\pi}\left(\frac{1}{3}+\log\frac{\mu^2}{-p^2}\right)\left[\frac{1}{N} Q_2^\text{SLL} - Q_1^\text{SLL}\right] \\
&\qquad+
\frac{\alpha_s}{4\pi}\left[\frac{2}{N} Q_4^\text{SLL} -2 Q_3^\text{SLL}\right]
\end{split}
\end{align}}%

 We remark that for the matching performed in this paper only QCD 
corrections to the matrix elements $\langle Q_2^\text{VLL}\rangle$, 
$\langle Q_2^\text{VLR}\rangle$, $\langle Q_2^\text{SLR}\rangle$ and
$\langle Q_2^\text{SLL}\rangle$ are required. The QCD corrections to the 
matrix elements of the remaining operators would enter the NLO analysis if  the
exchanged
gauge bosons and scalars were coloured. They would also enter our analysis 
if we included $\ord(\alpha_s^2)$ corrections. We include these additional 
matrix elements for completeness.

\section{Results for the Wilson Coefficients}\label{Sec5}

In what follows we will list the general structure of the effective Hamiltonian in each case and  subsequently we will list our results for
the Wilson coefficients (we drop $h.c.$ in what follows).

\subsection{Colourless gauge boson}\label{CLGB}

\begin{align}\begin{split}
 \mathcal{H}_\text{eff}^{\rm gauge} =&
\frac{\left(\Delta_L^{ub}(A)\right)^\star\Delta_L^{cd}(A)}{M_A^2}\left[C_1^\text{VLL}(\mu)Q_1^\text{VLL} +C_2^\text{VLL}(\mu)Q_2^\text{
VLL}  \right]\\
&+\frac{\left(\Delta_L^{ub}(A)\right)^\star\Delta_R^{cd} (A))^2 } {M_A^2 }
\left[C_1^\text{VLR}(\mu)Q_1^\text{VLR} +C_2^\text{VLR}(\mu)Q_2^\text{VLR} \right]
+L\leftrightarrow R\,.
\end{split}
\end{align}

We find for an arbitrary 
number of colours $N$
{\allowdisplaybreaks
\begin{align}
C_1^\text{VLL}(\mu) & =
\frac{\alpha_s}{4\pi}\left(-3\log\frac{M_A^2}{\mu^2}+\frac{11}{2}\right)\,,\\
 C_2^\text{VLL}(\mu) & =
1+\frac{\alpha_s}{4\pi}\left(\frac{3}{N}\log\frac{M_A^2}{\mu^2}-\frac{11}{2N}\right)=1+\frac{\alpha_s}{4\pi}
\left(\log\frac{M_A^2}{\mu^2}-\frac{11}{6}\right)\,,\\
C_1^\text{VLR}(\mu) &=\frac{\alpha_s}{4\pi}\left(3\log\frac{M_A^2}{\mu^2}+\frac{1}{2}\right)\,,\\
C_2^\text{VLR}(\mu) &=1+ \frac{\alpha_s}{4\pi}\left(-\frac{3} {N}\log\frac{M_A^2}{\mu^2}-\frac{1}{2N}\right) = 1+
\frac{\alpha_s}{4\pi}\left(-\log\frac{M_A^2}{\mu^2}-\frac{1}{6}\right)\,.
\end{align}}%

\subsection{Colourless scalar}\label{CLS}

\begin{align}\begin{split}
 \mathcal{H}_\text{eff}^{\rm scalar} =&
- \frac{\left(\Delta_L^{ub}(A)\right)^\star\Delta_L^{cd}(A)}{M_H^2}\left[C_1^\text{SLL}(\mu)Q_1^\text{SLL} +C_2^\text{SLL}(\mu)Q_2^\text{
SLL}  \right]\\
&
- \frac{\left(\Delta_L^{ub}(A)\right)^\star\Delta_L^{cd}(A)}{M_H^2}
\left[C_3^\text{SLL}(\mu)Q_3^\text{SLL} +C_4^\text{SLL}(\mu)Q_4^\text{SLL} \right]\\
&-\frac{\left(\Delta_L^{ub}(H)\right)^\star\Delta_R^{cd}(H)}{
M_H^2} \left [ C_1^\text{SLR}(\mu) Q_1^\text{SLR} +C_2^\text{SLR}(\mu) Q_2^\text{SLR} \right] +L\leftrightarrow R\end{split}
\end{align}
 We find for an arbitrary number of colours $N$
{\allowdisplaybreaks
\begin{align}
 C_1^\text{SLR}(\mu)& =3\frac{\alpha_s}{4\pi}\,,\\
C_2^\text{SLR}(\mu) &=  1-\frac{\alpha_s}{4\pi}\frac{3}{N}=  1-\frac{\alpha_s}{4\pi}\,,\\
C_1^\text{SLL}(\mu)&= 0\,,\\
C_2^\text{SLL}(\mu) &=  1\,,\\
C_3^\text{SLL}(\mu) &=\frac{\alpha_s}{4\pi}\left(-\frac{1}{2}\log\frac{M_H^2}{\mu^2}+\frac{3}{4}
\right)\,,\\
C_4^\text{SLL}(\mu)&=\frac{\alpha_s}{4\pi}\left(\frac{1}{2N}\log\frac{M_H^2}{\mu^2}-\frac{3}{4N}
\right) = \frac{\alpha_s}{4\pi}\left(\frac{1}{6}\log\frac{M_H^2}{\mu^2}-\frac{1}{4}
\right)\,.
\end{align}}%

We emphasize that the Wilson coefficients $C_i^a$ of the ``mirror'' operators ($P_L\leftrightarrow P_R$) as defined by us are equal to the
ones presented above.
 The formulae presented in this section  are the main results of our paper.

\section{Master Formulae for NLO Wilson Coefficient functions}\label{MasterNLO}
With these results at hand and the known one-loop and two-loop anomalous 
dimension matrices that we have listed in Section~\ref{Sec2} we can complete 
the NLO renormalization group analysis which can give us the Wilson 
coefficients at the low energy scales at which hadronic matrix elements 
are evaluated. In the case of the $B$ decays the final result for 
the Wilson coefficients in each of the four sectors of operators with no mixing 
between different sectors can be written as follows:
\be\label{RGevolution}
\vec{C}(\mu_b)=\hat{U}(\mu_b,\,\mu_{\rm in})\vec{C}(\mu_{\rm in}),
\ee
where $\hat{U}$ are evolution matrices and $\vec{C}$
column vectors.
For each VLL (VRR), VLR (VRL), SLR (SRL) sector $\hat{U}$ is a $2\times 2$ 
matrix, while it 
is a $4\times 4$ matrix in the  SLL (SRR) sector.
Similarly 
for each VLL (VRR), VLR (VRL), SLR (SRL) sector  $\vec{C}$ is a two-dimensional column vector, while it 
is a four-dimensional one in the  SLL (SRR) sector.

General formulae for the evolution matrix at the NLO level 
have been derived in \cite{Buras:1991jm}. In order to make our paper self-contained we 
recall these formulae in what follows:
\begin{equation}\label{u0jj}
\hat{U}(\mu_b,\mu_{\rm in})=
\left(\mathds{1}+\frac{\alpha_s(\mu_b)}{4\pi} \hat{J}\right) \hat{U}^{(0)}(\mu_b,\mu_{\rm in})
\left(\mathds{1}-\frac{\alpha_s(\mu_{\rm in})}{4\pi} \hat{J}\right).
\end{equation}
Here $\hat{U}^{(0)}$ is the evolution matrix in leading logarithmic approximation
and the matrix $\hat{J}$ expresses the NLO corrections to this
evolution. We have
\begin{equation}\label{u0vd} 
\hat{U}^{(0)}(\mu_b,\mu_{\rm in})= \hat{V}
\left({\left[\frac{\alpha_s(\mu_{\rm in})}{\alpha_s(\mu_b)}\right]}^{\frac{\vec\gamma^{(0)}}{2\beta_0}}
   \right)_D \hat{V}^{-1} \,,  \end{equation}
where $\hat{V}$ diagonalizes ${\hat{\gamma}^{(0)\top}}$
\begin{equation}\label{ga0d} 
\hat{\gamma}^{(0)}_D=\hat{V}^{-1} {\hat{\gamma}^{(0)\top}} \hat{V}  \end{equation}
and $\vec\gamma^{(0)}$ is the vector containing the diagonal elements of
the diagonal matrix $\hat{\gamma}^{(0)}_D$.\\
If we define
\begin{equation}\label{gvg1} 
\hat{G}=\hat{V}^{-1} {\hat{\gamma}^{(1)\top}} \hat{V}   \end{equation}
and a matrix $\hat{H}$ whose elements are
\begin{equation}\label{sij} 
H_{ij}=\delta_{ij}\gamma^{(0)}_i\frac{\beta_1}{2\beta^2_0}-
    \frac{G_{ij}}{2\beta_0+\gamma^{(0)}_i-\gamma^{(0)}_j}\,,  \end{equation}
the matrix $\hat{J}$ is given by
\begin{equation}\label{jvs} 
\hat J=\hat{V}\hat{H} {\hat{V}}^{-1}   \,.
\end{equation}

We next write our results for the Wilson coefficients at the matching scale 
in a general form as follows
\begin{align}
 \vec{C}(\mu_{\rm in}) = \vec{C}_0 - \frac{\alpha_s(\mu_{\rm in})}{4\pi}\vec{C}_1 \,.
\end{align}
Finally combining these initial values 
with the evolution matrix ({\ref{u0jj}) we obtain
\begin{equation}\label{cjua} \vec C(\mu_b)=
\left(\mathds{1}+\frac{\alpha_s(\mu_b)}{4\pi} \hat{J}\right) \hat{U}^{(0)}(\mu_b,\mu_{\rm in})
\left(\mathds{1}-\frac{\alpha_s(\mu_{\rm in})}{4\pi}\left(\vec{C}_1+\hat{J}\vec{C}_0\right)\right).
\end{equation}

We recall that when using this evolution down to low energy scales one has to 
insert the correct number of effective flavours. As this procedure is by now 
standard, we refer to Section IIIE in \cite{Buchalla:1995vs} for details.

We end this section by recalling certain features of the fundamental formula~(\ref{cjua}):
\begin{itemize}
\item
The renormalization scheme dependence of the matrix $\hat{J}$ on the left-hand 
side of the LO evolution matrix is cancelled by the one of hadronic matrix 
elements. 
\item 
This scheme dependence on the right-hand side of the LO evolution matrix is 
cancelled by the one of $\vec{C}_1$ calculated by us: 
$\vec{C}_1+\hat{J}\vec{C}_0$ is renormalization scheme independent.
\item
The dependence on the precise choice of the scale $\mu_b$ in the  
evolution matrix is cancelled 
by the one present in the hadronic matrix elements. In the case of $\Delta F=2$ 
transitions in the SM, in which only a single operator is present this allows 
to introduce renormalization scheme and renormalization scale invariant 
parameters like $\hat{B}_K$.
\item
The dependence on the precise choice of the scale $\mu_{\rm in}$ in the 
evolution matrix is cancelled by the logarithmic terms in $\vec{C}_1$ that 
we calculated in this paper.
\end{itemize}

We will now look in more details at the last issue.

\boldmath
\section{Renormalization Scale Dependence}\label{Sec6}
\unboldmath

One of the main virtues of our calculation of $\mathcal{O}(\alpha_s)$ corrections to Wilson coefficients at the high energy matching scale
$\mu_{\rm in}$ is the cancellation of the $\mu_{\rm in}$ dependence of the renormalization group evolution matrix by the $\mu_{\rm in}$ dependence of the Wilson
coefficients in question. This cancellation requires particular values of the coefficients of the $\log(M^2/\mu_{\rm in}^2)$ in
$C_i(\mu_{\rm in})$ where $M$ stands for the mass of a heavy gauge boson or heavy scalar involved. As this cancellation constitutes an important test of our results
it is useful to derive a general condition on the coefficients of 
$\log(M^2/\mu_{\rm in}^2)$ in $C_i(\mu_{\rm in})$.

To this end let us look at the evolution matrix in Eq.~(\ref{RGevolution}).
Expanding this matrix
 around the two fixed scales $m_b$ and $M$ keeping
only the logarithmic terms one obtains
\begin{align}\label{equ:hatU}
 \hat U(\mu_b,\,\mu_{\rm in}) =
\left(\mathds{1}+\frac{\alpha_s(\mu_b)}{4\pi}\frac{\hat\gamma^{(0)\top}}{2}\log\frac{\mu_b^2}{m_b^2}\right)\hat{U}(m_b,\,M)
\left(\mathds{1}+\frac{\alpha_s(\mu_\text{in})}{4\pi}\frac{\hat\gamma^{(0)\top}}{2}\log\frac{M^2}{\mu_{\rm in}^2}\right)\,,
\end{align}
where $\hat\gamma^{(0)}$ is the coefficient of $\alpha_s$ in the one loop anomalous dimension matrix that describes the mixing of operators.
The $\hat\gamma^{(0)}$ matrices for VLL (VRR), VLR (VRL), SLR (SRL) and SLL (SRR) 
sectors  have been collected 
in Section~\ref{Sec2}.
Note that it is $\hat\gamma^{(0)\top}$ and not $\hat\gamma^{(0)}$ that enters Eq.~(\ref{equ:hatU}).
Moreover, in the study of the $\mu_{\rm in}$ dependence in the case of the 
scalar exchange one has to take into account that in this case 
the $m^2(\mu_{\rm in})$ dependence is hidden in
the coefficients $(\Delta_{L/R}^{ij}(H))^\star\Delta_{L/R}^{kl}(H)$.

Considering then the cases of colourless gauge bosons and scalars we find that the following quantities should be $\mu_{\rm in}$ -- independent:
\begin{subequations}\label{equ:R}
\begin{align}
 R^\text{gauge} & = \hat U(\mu_b,\,\mu_{\rm in})\vec{C}(\mu_{\rm in})\,,\\
 R^\text{scalar} & = \hat U(\mu_b,\,\mu_{\rm in})\vec{C}(\mu_{\rm in})m^2(\mu_{\rm in})\,.
 \end{align}
\end{subequations}
For each VLL (VRR), VLR (VRL), SLR (SRL) sector  $\vec{C}(\mu)$ is a two-dimensional column vector, while it 
is four-dimensional for the  SLL (SRR) sector.

We write next in each case
\begin{align}
 \vec{C}(\mu_{\rm in}) = \vec{C}_0 - \frac{\alpha_s(\mu_{\rm in})}{4\pi}\vec{K} \log\frac{M^2}{\mu_{\rm in}^2}\,,
\end{align}
where we suppressed $\mu_{\rm in}$ independent $\mathcal{O}(\alpha_s)$ terms and
\begin{align}
 m^2(\mu_{\rm in}) = m^2(M)\left(1+ \frac{\alpha_s(\mu_\text{in})}{4\pi}\gamma^{(0)}_m \log\frac{M^2}{\mu_{\rm in}^2}\right)\,,
\end{align}
with $\gamma^{(0)}_m$ governing the scale dependence of quark masses in QCD.

Imposing Eq.~(\ref{equ:R}), the conditions for $\vec{K}$ to ensure 
$\mu_{\rm in}$ independence of resulting amplitudes in these two cases read
\begin{subequations}\label{equ:vecK}
\begin{align}
 \vec{K}^\text{gauge}& = \frac{\hat\gamma^{(0)\top}}{2}\vec{C}_0\,,\\
\vec{K}^\text{scalar} & = \frac{\hat\gamma^{(0)\top}}{2}\vec{C}_0+\gamma^{(0)}_m\vec{C}_0\,.
\end{align}
\end{subequations}
Thus the coefficients of logarithms in $\vec{C}(\mu_{\rm in})$ can be found without the calculation of the loop diagrams in
Fig.~\ref{fig:fullgauge}-\ref{fig:eff} but formulae in Eq.~(\ref{equ:vecK})  serve as a useful check of our results for
logarithmic terms. These
terms are renormalization scheme independent and while cancelling 
the $\mu_{\rm in}$ dependence of $\hat U(\mu_b,\,\mu_{\rm in})$ in perturbation theory
cannot remove its renormalization scheme dependence at the NLO level. To this end as discussed in Section~\ref{MasterNLO},
the $\mathcal{O}(\alpha_s)$ non-logarithmic terms have to
be calculated which constitutes the main new result of our paper. 

Inserting  the formulae for the one-loop anomalous dimension matrices and 
$\gamma^{(0)}_m$, that we listed in Section~\ref{Sec2}, into Eq.~(\ref{equ:vecK}) one can verify that  
the resulting coefficients $\vec{K}^\text{gauge}$ 
and $\vec{K}^\text{scalar}$ equal the coefficients of the 
logarithmic terms calculated by us. This implies that the inclusion 
of $\ord(\alpha_s)$ corrections in question remove the unphysical dependence on 
the precise value of the matching scale.

The manner in which the
$\mu_{\rm in}$-dependence is removed resembles the one which we encountered 
in the case of $\Delta F=2$ transitions \cite{Buras:2012fs}:
\begin{itemize}
\item
 In the gauge boson case the  $\mu_{\rm in}$-dependence of 
$\hat U(\mu_b,\,\mu_{\rm in})$ can only be cancelled by the corrections 
calculated by us. 
\item
The case of scalar exchange with SLR couplings is quite different.
Here the $\mu_{\rm in}$-dependence of $\hat U(\mu_b,\,\mu_{\rm in})$ is totally 
cancelled by the one of the $ m^2(\mu_{\rm in})$ so that even without 
our corrections the amplitudes are $\mu_{\rm in}$ independent. 
The role of our calculation  in this case is 
then the removal of the renormalization scheme dependence.
\item
In the case of SLL operators both the running quark masses and the 
corrections calculated by us are required for the removal of the 
unphysical matching scale dependence present in LO.
 \end{itemize}

\section{Numerical Analysis}\label{Sec6a}
We will now compute the size of unphysical $\mu_{\rm in}$-dependence 
present in the LO expressions and we will demonstrate their 
reduction after the inclusion of $\ord(\alpha_s)$ corrections. 

Compared to our analysis of $\Delta F=2$ processes in \cite{Buras:2012fs} a numerical analysis of left-over unphysical scale dependences in
the decay
amplitudes is 
complicated by three facts:
\begin{itemize}
\item
The analogues of $P_i^a$ factors \cite{Buras:2001ra} that summarize the renormalization group effects between 
high energy and low energy scales and include also the values of hadronic 
matrix elements do not exist in the literature for decays discussed here 
although obviously complete NLO analyses for the VLL case are known.
\item
The hadronic matrix elements of new operators contain larger theoretical 
uncertainties than in the case of $\Delta F=2$ decays.
\item
The increased number of operators relative to the ones in the case of 
$\Delta F=2$ transitions makes the analysis in questions very involved.
\end{itemize}

Leaving a detailed analysis in a concrete model for the future, we 
nevertheless illustrate the size of unphysical $\mu_{\rm in}$-dependence 
present in the LO expressions  in the  
sectors VLL, VLR and SLL. In the SLR sector our corrections have no impact on 
the cancellation of the $\mu_{\rm in}$ dependence as 
we discussed above.

We will solve the problem of poorly known hadronic elements in the 
following manner. We will go to the operator basis in which 
the one-loop anomalous dimension matrices are diagonal. At LO  
 in each of the VLL, VLR and SLR sectors there are then two-operators that do not mix with 
each other and with other operators of other sectors.  There are four such 
operators in the SLL sector. This property 
remains true at NLO only in the VLL sector as only in this sector 
the one-loop and two-loop anomalous dimension matrices are diagonalized 
by the same matrix $\hat{V}$. For VLR and also scalar sectors one
has to use the general formulae presented in Section~\ref{MasterNLO}.

Yet, as we have seen in the previous section in order to study the 
cancellation of the $\mu_{\rm in}$-dependence present in the LO 
expressions it is sufficient to study the evolution matrix at the 
leading order and keep only the leading logarithms in $\vec{C}^{(1)}$. 
Diagonalizing then the matrices $\hat{\gamma}^{(0)}$  in all four
sectors we can factor out the hadronic matrix element in each 
case so that the issue of the study of the $\mu_{\rm in}$-dependence 
as expected can be investigated transparently without any hadronic uncertainties.

 Denoting by $C^a_{\pm}$ the Wilson coefficients of the operators 
$Q^a_{\pm}$ corresponding to the eigenvalues 
 of the two anomalous dimension 
 matrices $a= {\rm VLL,VLR}$ and normalizing $C^a_{\pm}$ so that they 
are equal unity at the matching scale at LO we find:

{\bf VLL Case}

\be
Q_+^{\rm VLL}=\frac{Q_2^{\rm VLL}+Q_1^{\rm VLL}}{2}, \quad 
C_+^{\rm VLL}=C_2^{\rm VLL}+C_1^{\rm VLL},\quad \gamma^{(0)+}_{\rm VLL} =-\frac{6}{N}+6,
\ee

\be
Q_-^{\rm VLL}=\frac{Q_2^{\rm VLL}-Q_1^{\rm VLL}}{2}, \quad 
C_-^{\rm VLL}=C_2^{\rm VLL}-C_1^{\rm VLL},\quad \gamma^{(0)-}_{\rm VLL}=-\frac{6}{N}-6.
\ee

{\bf VLR Case}

\be
Q_+^{\rm VLR}=Q_2^{\rm VLR}-\frac{Q_1^{\rm VLR}}{N}, \quad 
C_+^{\rm VLR}=C_2^{\rm VLR},\quad \gamma^{(0)+}_{\rm VLR}=\frac{6}{N},
\ee

\be
Q_-^{\rm VLR}=\frac{Q_1^{\rm VLR}}{N}, \quad 
C_-^{\rm VLR}=C_2^{\rm VLR}+N C_1^{\rm VLR},\quad \gamma^{(0)-}_{\rm VLR}=\frac{6}{N}-6N.
\ee

Here $\gamma_a^{(0)\pm}$ denote the anomalous dimensions of the operators 
$Q^a_{\pm}$. Then the  quantities ($a={\rm VLL,VLR}$)
 \be \label{Ri1}
 R_a^\pm= \left[\frac{\alpha_s(\mu_{\rm in})}
{\alpha_s(M)}\right]^{\frac{\gamma_a^{(0)\pm}}{2\beta_0}}
 \left(1+ \frac{\alpha_s(\mu_{\rm in})}{4\pi} \left[ - K_a^\pm \log\frac{M^2}{\mu_{\rm in}^2}+ r_a^\pm\right]\right)\,,
     \end{equation}
where $r_a^\pm$ denote the non-logarithmic $\ord(\alpha_s)$ corrections in the new basis
should be  $\mu_\text{in}$ independent at $\ord(\alpha_s)$. Here $M$ is 
the mass of  the exchanged gauge boson that we will set to 
$1\tev$ in our numerical calculations.

 We should emphasize that in addition to $\ord(\alpha_s)$ corrections proportional 
to $r_a^\pm$ there are also corrections at this order that come from the 
evolution matrix. We have denoted them by $\hat{J}$ in Section~\ref{MasterNLO}. We do 
not include them in our analysis in order to exhibit the size of the 
corrections calculated here but they have to be included in any phenomenological analysis in order to obtain renormalization scheme
independent results.

From our discussion of the previous section and the diagonalization 
performed here we find that
\be
K_a^\pm=\frac{\gamma_a^{(0)\pm}}{2}, \qquad a={\rm VLL,VLR}
\ee
\begin{subequations}\label{equ:rVLLVLR}
\begin{align}
 & r_\text{VLL}^+ = \frac{11(N-1)}{2N}= \frac{11}{3}\,,& &r_\text{VLL}^- = -\frac{11(N+1)}{2N} = -\frac{22}{3}\,,\\
& r_\text{VLR}^+ = -\frac{1}{2N} = -\frac{1}{6}\,,&  &r_\text{VLR}^- = \frac{N^2-1}{2N} =\frac{4}{3}\,,
\end{align}
\end{subequations}
and indeed then the $R_a^\pm$ should be equal unity after the inclusion 
of only the $\ord(\alpha_s)$ logarithmic corrections $K_a^\pm$ calculated here. 
The departure from unity at NLO signals the presence of $r_a^\pm$ terms.

\begin{figure}[!tb]
 \centering

\includegraphics[width=.48\linewidth]{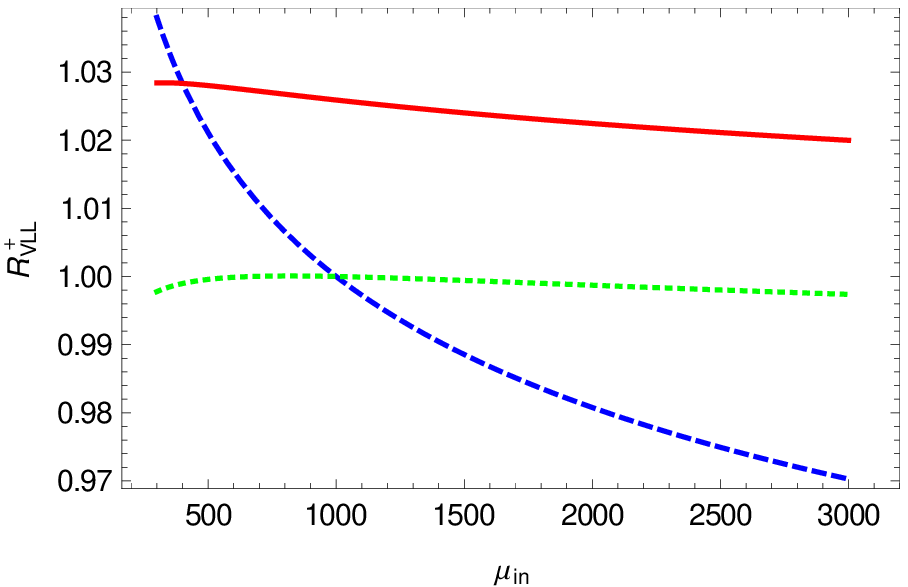}
\hspace{0.2cm}
\includegraphics[width=.48\linewidth]{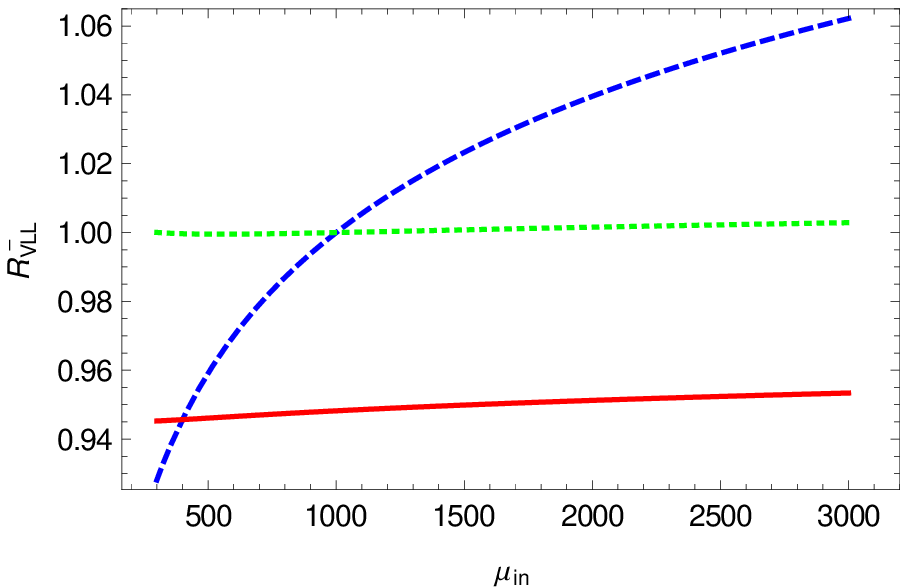}

\includegraphics[width=.48\linewidth]{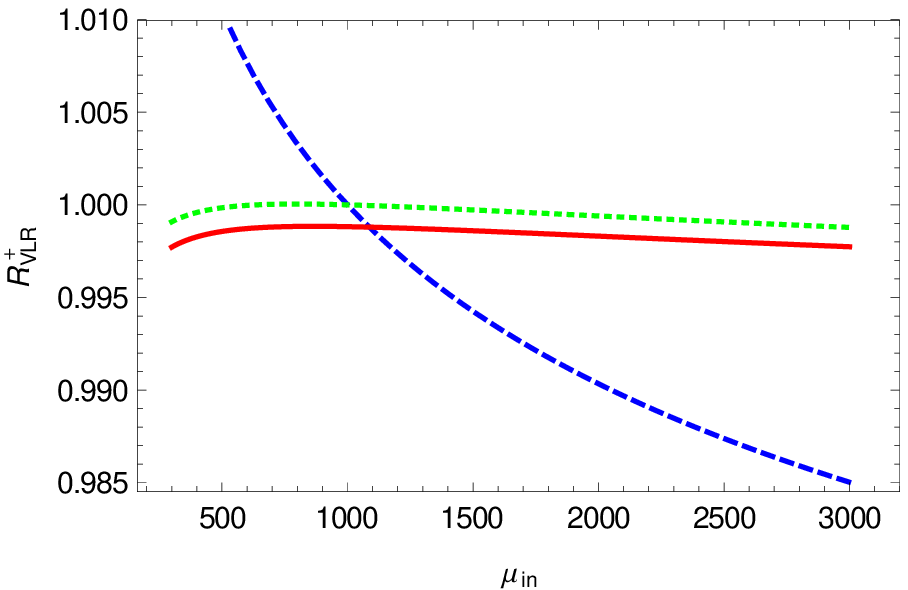}
\hspace{0.2cm}
\includegraphics[width=.48\linewidth]{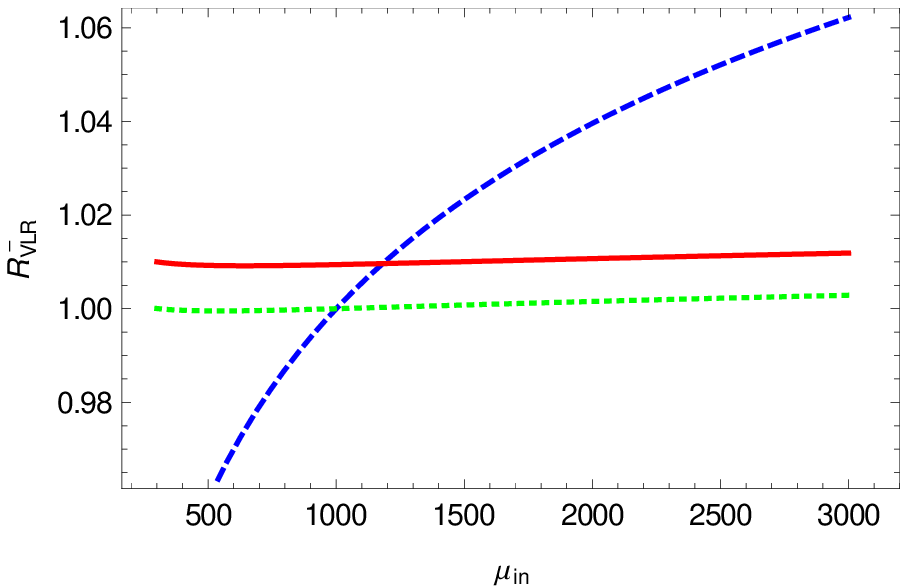}

\caption{\it The quantities $R_a^\pm$ ($a =$ VLL, VLR) defined in Eq.~(\ref{Ri1}) as a function of $\mu_\text{in}$ for $M =
1~$TeV. The LO result (removing contribution proportional to $K^\pm_a$ and $r_a^\pm$) is shown by the dashed blue line, the dotted green
line is the NLO result including only logarithmic $\ord(\alpha_s)$ corrections $K^\pm_a$ and the solid red line shows the NLO result
including both $K^\pm_a$ and $r_a^\pm$.}\label{fig:plots}
\end{figure}

 In Fig.~\ref{fig:plots} we plot $R_a^\pm$  for $a={\rm VLL,VLR}$ as functions of the matching scales setting as 
an example the masses of  gauge bosons to $1\tev$ for three cases: Without the $\ord(\alpha_s)$ corrections in the Wilson
coefficients (dashed blue line), with the contribution proportional to $K_a^\pm$ (dotted green line) and including both 
logarithmic $K_a^\pm$ and non-logarithmic $r_a^\pm$ corrections  (solid red line).  While the dashed blue lines
exhibit a significant $\mu_\text{in}$ dependence, the dotted green  lines and solid red lines  stay
nearly constant  over the considered $\mu_\text{in}$ range. The dotted green lines that include only logarithmic $\ord(\alpha_s)$
corrections are
equal to 1 at $\mu_{\rm in}=1\tev$ as expected. For the solid red lines a small  shift relative to the dotted green ones occurs which is due
to the non-logarithmic
corrections $r_a^\pm$.  Only in the VLL sector  which has larger $r_a^\pm$ than the VLR sector  a
slight $\mu_\text{in}$ dependence occurs. This dependence can only be cancelled 
by NNLO corrections.

{\bf SLL Case}

The diagonalization in the SLL case is a bit more involved since four operators are present. Furthermore, in the scalar case we also have to
take into account the $\mu_\text{in}$ dependence of the quark masses (see Eq.~(\ref{equ:R})). Thus instead of
Eq.~(\ref{Ri1}) the
following quantities  
\begin{align}
  R_\text{SLL}^j= \left[\frac{\alpha_s(\mu_{\rm in})}
{\alpha_s(M)}\right]^{\frac{\gamma_\text{SLL}^{(0)j}}{2\beta_0}}
 \left(1+ \frac{\alpha_s(\mu_{\rm in})}{4\pi} \left[ - K_\text{SLL}^j \log\frac{M^2}{\mu_{\rm in}^2}+
r_\text{SLL}^j\right]\right)\frac{m^2(\mu_\text{in})}{m^2(M)}\,,\quad j = \pm\pm,\pm
\end{align}
should be $\mu_\text{in}$ independent at $\ord(\alpha_s)$. For simplicity we set in the following $N = 3$ in order to shorten the
formulae.
We find
\begin{align}
 &K_\text{SLL}^j = \frac{\gamma_\text{SLL}^{(0)j}}{2}+\gamma_m^{(0)}\,,\quad j = \pm\pm,\pm\,,
\end{align}
with
\begin{align}\label{equ:gSLL}
 & \gamma_\text{SLL}^{(0)\pm\pm} =  \frac{2}{3}(1\pm\sqrt{241})\,,&\quad&\gamma_\text{SLL}^{(0)\pm} =  \frac{2}{3}(-17\pm\sqrt{241})\,.
\end{align}
The non-logarithmic corrections in the basis in which $\gamma^{(0)}_\text{SLL}$ is diagonal and LO Wilson coefficients $C_\text{SLL}^j$
($j = \pm\pm,\pm$) are normalized to 1 read
\begin{align}
& r_\text{SLL}^{\pm\pm} = \frac{1}{2}(25 \pm \sqrt{241})\,,& & r_\text{SLL}^\pm = \frac{1}{2}(7\pm\sqrt{241})\,.
\end{align}
In the Appendix~\ref{app} we list the Wilson coefficients and operators in this new basis.

\begin{figure}[!tb]
 \centering
\includegraphics[width=.48\linewidth]{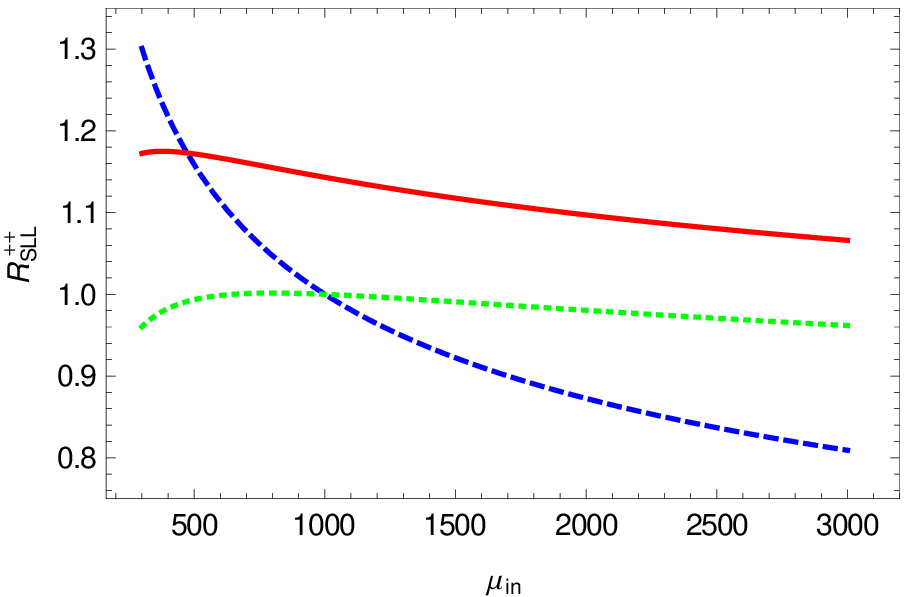}
\hspace{0.2cm}
\includegraphics[width=.48\linewidth]{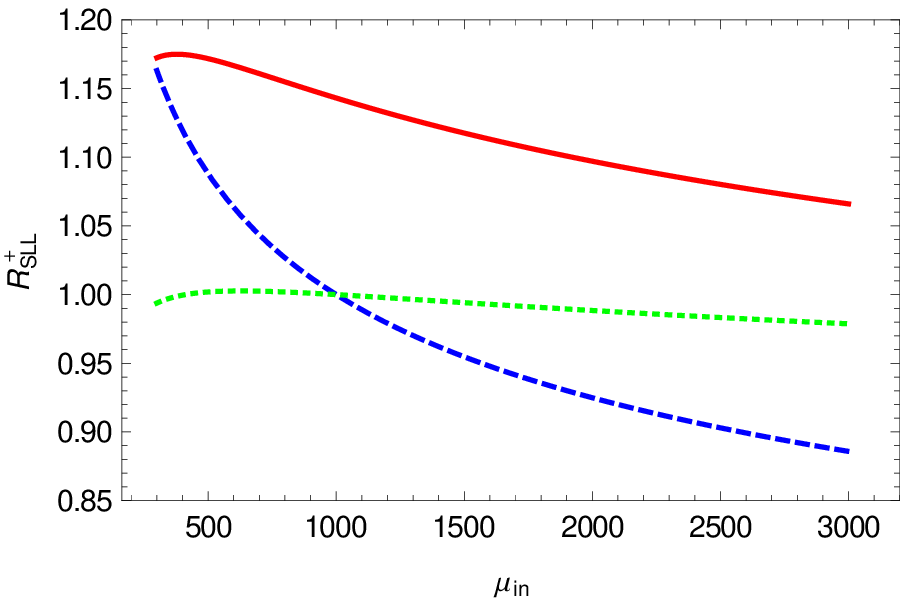}

\includegraphics[width=.48\linewidth]{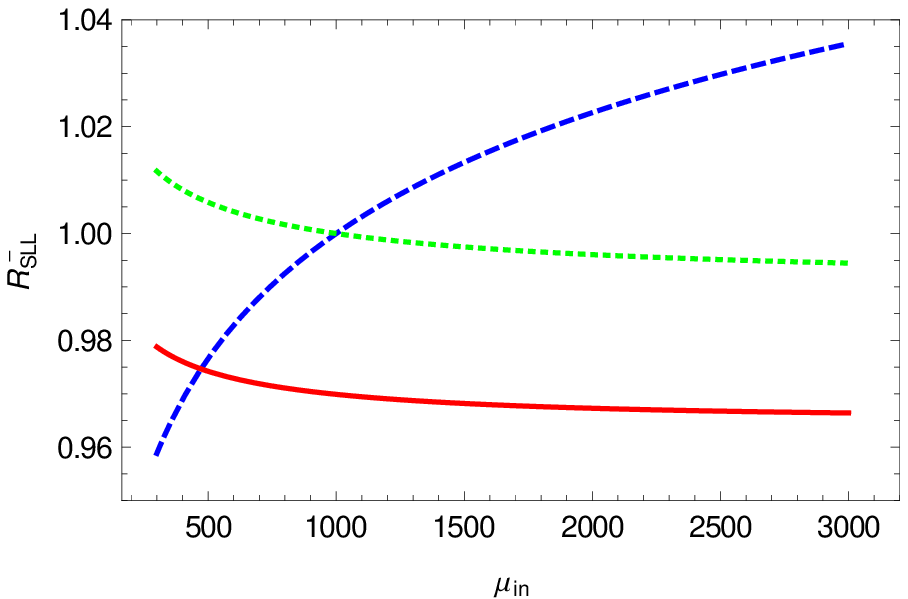}
\hspace{0.2cm}
\includegraphics[width=.48\linewidth]{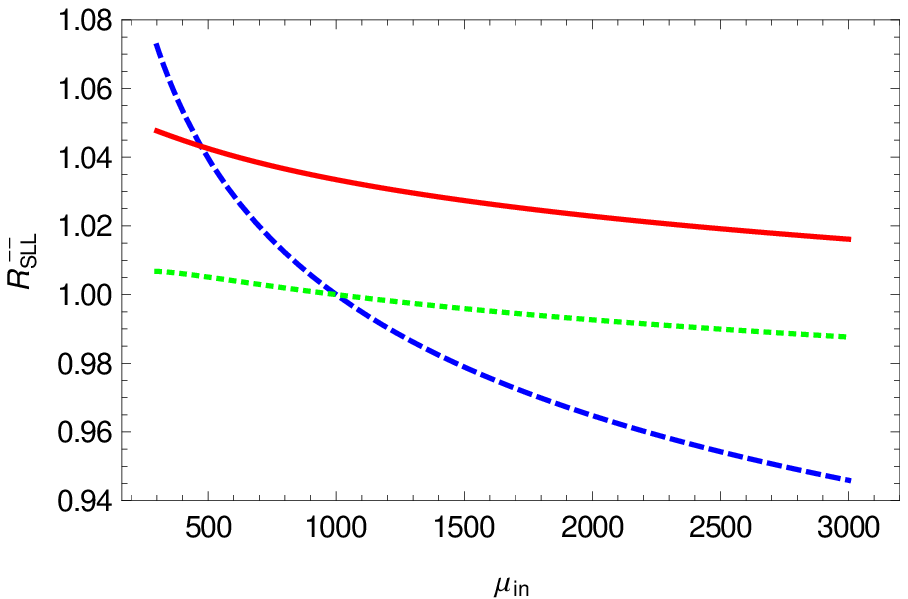}
\caption{\it The quantities $R_\text{SLL}^j$ ($j = \pm\pm,\,\pm)$ as a function of $\mu_\text{in}$ for $M =
1~$TeV. The LO result (removing contribution proportional to $K^j_\text{SLL}$ and $r_\text{SLL}^j$) is shown by the dashed blue line, the
dotted green
line is the NLO result including only logarithmic $\ord(\alpha_s)$ corrections $K^j_\text{SLL}$ and the solid red line shows the NLO result
including both $K^j_\text{SLL}$ and $r_\text{SLL}^j$. }\label{fig:plots2}
\end{figure}

In Fig.~\ref{fig:plots2} we show the results for the SLL sector, again for the three different cases: LO (dashed blue), NLO with only
logarithmic corrections (dotted green) and NLO with both logarithmic and non-logarithmic corrections (solid red).
 As expected from the values of $K^j_{\rm SLL}$ the dashed blue lines show 
a very strong $\mu_\text{in}$ dependence in the case of $++$ and $+$. The inclusion of the logarithmic 
$\ord(\alpha_s)$ terms calculated by us practically removes this dependence 
(dotted green lines). However, the large size of the non-logarithmic terms 
 $r_\text{SLL}^{++/+}$ implies significant left-over $\mu_\text{in}$ dependence in 
these two cases at the NLO level (solid red lines) that can only be removed by NNLO corrections.

{\bf SLR case}

At last we discuss the special case SLR where no logarithms appear in the matching conditions. Diagonalizing
$(\gamma^{(0)\text{SLR}})^\top$ we get
\begin{align}
 &Q_+^\text{SLR} = Q_2^\text{SLR} - N Q_1^\text{SLR}\,,&& C_+^\text{SLR} = -\frac{1}{N}C_1^\text{SLR}\,,&&
\gamma_\text{SLR}^{(0)+} = \frac{6}{N}\,,\\
&Q_-^\text{SLR} = Q_2^\text{SLR}\,,&& C_-^\text{SLR} = \frac{1}{N}C_1^\text{SLR} + C_2^\text{SLR}\,,&& \gamma_\text{SLR}^{(0)-} =
-6 N + \frac{6}{N}\,.\label{equ:SLR-}
\end{align}
We can read off that the coefficient $C_+^\text{SLR}$ cannot be unity in leading order. Only $C_-^\text{SLR}$ can be normalized to 1 and
its $\mu_\text{in}$ dependence related to the anomalous dimension of $Q_-^\text{SLR}$ is exactly cancelled by  the
$\mu_\text{in}$ dependece of quark masses. This means simply  that $\gamma_\text{SLR}^{(0]-}= -2\gamma_m^{(0)}$ as seen explicitely in
Eq.~(\ref{equ:SLR-}).

\section{Summary}\label{Sec7}

In various extensions of the Standard Model (SM) 
tree level non-leptonic decays of 
hadrons receive contributions from new heavy gauge bosons and scalars. 
Prominent examples are the right-handed $W^\prime$ bosons in left-right symmetric models and charged Higgs ($H^\pm$) particles in models
with extended scalar sector like  two Higgs doublet models and supersymmetric models. 
Of particular interest are the decays in which the contributing local 
operators involve four different flavours so that penguin operators 
cannot contribute and the decays are simpler to analyse theoretically.
Anticipating the important role of such decays at the LHCb, KEKB and Super-B in 
Rome and having in mind future improved lattice calculations combined with the QCD factorization approach, we have 
 completed 
the existing NLO QCD calculations of these processes by calculating
$\mathcal{O}(\alpha_s)$
corrections to matching conditions for the Wilson coefficients of all 
contributing operators in the NDR-$\overline{\text{MS}}$ scheme in any extension of the SM.
The main results of our paper can be found in Section~\ref{Sec5}.

Our calculation
allowed to reduce certain unphysical scale and renormalization scheme
dependences in the existing NLO calculations. Our results can also be 
applied to models with tree-level heavy neutral gauge boson and scalar 
exchanges in $\Delta F = 1$ decays and constitute an important part of NLO analyses of those non-leptonic
 decays to which 
also penguin operators contribute.

For completeness we have collected all the relevant formulae necessary to 
perform the  full NLO renormalization group analysis that would require 
the evaluation of the hadronic matrix elements in the 
the NDR-$\overline{\text{MS}}$ scheme.
 They can be found in 
Section~\ref{MasterNLO} with all ingredients given in Sections~\ref{Sec2} and 
\ref{Sec5}.

We are aware of the fact that present hadronic uncertainties in the decays considered are significantly larger than corrections calculated
by us. However, one should recall that twenty years ago similar comments were made in connection with NLO QCD corrections within the SM
\cite{Buras:2011we}. During the last decade after significant progress in lattice calculation has been made, the results of the 1990's
constitute an important part of any phenomenological analysis of weak decays in the SM. We expect that this will be the case of our results
in due time.

\subsection*{Acknowledgements}
This research was done in the context of the ERC Advanced Grant project ``FLAVOUR''(267104) and was partially supported by the DFG cluster
of excellence 
``Origin and Structure of the Universe''.

\appendix

\section{Change of basis: SLL case}\label{app}

For our numerics in Section~\ref{Sec6a} we had to diagonalize the anomalous dimension matrix in the SLL sector. In this case the situation
is a bit more
intricate as four operators are involved. The eigenvalues $\gamma_\text{SLL}^{(0)j}$ ($j = \pm\pm,\,\pm$) are given in
Eq.~(\ref{equ:gSLL}) for $N = 3$. For completeness we list here 
the Wilson coefficients and operators in the basis in which $\gamma^{(0)\text{SLL}}$ is diagonal and where LO Wilson coefficients
$C_\text{SLL}^j$ ($j = \pm\pm,\,\pm$)  are normalized to 1.  We recall, as found 
in Section~\ref{Sec5}, that in the original basis only the coefficient $C_2^{\rm SLL}$ is non-vanishing in the LO.
{\allowdisplaybreaks
\begin{subequations}
\begin{align}
 Q_{\pm\pm}^\text{SLL} = &\left(-\frac{1}{8}  \pm \frac{19}{8\sqrt{241}}\right)Q_1^\text{SLL} +\left(\frac{1}{4} \mp
\frac{4}{\sqrt{241}}\right)Q_2^\text{SLL}  + \left(\frac{1}{32}  \mp \frac{15}{32\sqrt{241}}\right)Q_3^\text{SLL}\nonumber\\
&  \pm 
\frac{1}{16\sqrt{241}}Q_4^\text{SLL}\,,\\
 C_{\pm\pm}^\text{SLL} =& \frac{1}{2}(15\pm\sqrt{241}) C_1^\text{SLL} + C_2^\text{SLL}+2(19\pm\sqrt{241}) C_3^\text{SLL} +
4(16 \pm\sqrt{241}) C_4^\text{SLL}\,,\nonumber\\
 Q_{\pm}^\text{SLL} = &\left(\frac{1}{8}  \mp \frac{1}{8\sqrt{241}}\right)Q_1^\text{SLL} +\left(\frac{1}{4} \mp
\frac{4}{\sqrt{241}}\right)Q_2^\text{SLL}  + \left(-\frac{1}{32}  \pm \frac{21}{32\sqrt{241}}\right)Q_3^\text{SLL}\nonumber\\
&  \mp 
\frac{5}{16\sqrt{241}}Q_4^\text{SLL}\,,\\
 C_{\pm}^\text{SLL} = &\frac{1}{10}(21\pm\sqrt{241}) C_1^\text{SLL} + C_2^\text{SLL}+\frac{2}{5}(1\pm\sqrt{241}) C_3^\text{SLL} +
\frac{4}{5}(-16 \mp \sqrt{241}) C_4^\text{SLL}\,.\nonumber
\end{align}
\end{subequations}}%


\bibliographystyle{JHEP}
\bibliography{LitBG}

\end{document}